\documentclass[prd,amsmath,amssymb,showpacs,superscriptaddress,nofootinbib,twocolumn]{revtex4-1}
\usepackage{graphicx}
\usepackage{epsfig,graphics,subfigure,psfrag,amsmath,amssymb}
\usepackage{lineno}
\usepackage{dcolumn}
\usepackage{bm}
\usepackage{overpic}
\usepackage{xspace}
\usepackage{xcolor}
\usepackage{rotating}
\usepackage{color}
\usepackage{multirow}
\usepackage{colortbl}
\usepackage{epstopdf}
\usepackage{float}
\usepackage{makecell}
\usepackage[colorlinks,linkcolor=blue,anchorcolor=blue,citecolor=blue]{hyperref}

\newcommand{\BESIII}{BES\uppercase\expandafter{\romannumeral3}\xspace}

\begin{document}
\title{\Large \boldmath \bf Study of $\eta\rightarrow\pi^+\pi^-l^+l^-$}
\author{
\begin{small}
\begin{center}
M.~Ablikim$^{1}$, M.~N.~Achasov$^{4,c}$, P.~Adlarson$^{76}$, O.~Afedulidis$^{3}$, X.~C.~Ai$^{81}$, R.~Aliberti$^{35}$, A.~Amoroso$^{75A,75C}$, Q.~An$^{72,58,a}$, Y.~Bai$^{57}$, O.~Bakina$^{36}$
% ...
\\
\vspace{0.2cm}
(BESIII Collaboration)\\
\vspace{0.2cm} {\it
$^{1}$ Institute of High Energy Physics, Beijing 100049, People's Republic of China\\
% ...
$^{3}$ Bochum  Ruhr-University, D-44780 Bochum, Germany\\
$^{4}$ Budker Institute of Nuclear Physics SB RAS (BINP), Novosibirsk 630090, Russia\\
% ...
$^{35}$ Johannes Gutenberg University of Mainz, Johann-Joachim-Becher-Weg 45, D-55099 Mainz, Germany\\
$^{36}$ Joint Institute for Nuclear Research, 141980 Dubna, Moscow region, Russia\\
% ...
$^{57}$ Southeast University, Nanjing 211100, People's Republic of China\\
$^{58}$ State Key Laboratory of Particle Detection and Electronics, Beijing 100049, Hefei 230026, People's Republic of China\\
% ...
$^{72}$ University of Science and Technology of China, Hefei 230026, People's Republic of China\\
% ...
$^{75}$ University of Turin and INFN, (A)University of Turin, I-10125, Turin, Italy; (B)University of Eastern Piedmont, I-15121, Alessandria, Italy; (C)INFN, I-10125, Turin, Italy\\
$^{76}$ Uppsala University, Box 516, SE-75120 Uppsala, Sweden\\
% ...
$^{81}$ Zhengzhou University, Zhengzhou 450001, People's Republic of China\\

\vspace{0.2cm}
$^{a}$ Deceased\\
% ...
$^{c}$ Also at the Novosibirsk State University, Novosibirsk, 630090, Russia\\
% ...

}

\end{center}
\vspace{0.4cm}
\end{small}
}

\begin{abstract}
Using a sample of $(10087\pm44)\times10^{6}$ $J/\psi$ events accumulated with the BESIII detector, we analyze the decays $\eta\rightarrow\pi^+\pi^-l^+l^-$ ($l=e$ or $\mu$) via the process $J/\psi\rightarrow\gamma\eta$. 
The branching fraction of $\eta\rightarrow\pi^+\pi^-e^+e^-$ is measured to be $\mathcal{B}(\eta\rightarrow\pi^+\pi^-e^+e^-)=(3.07\pm0.12_{\rm{stat.}}\pm0.19_{\rm{syst.}}) \times10^{-4}$.
No signal events are observed for the $\eta\rightarrow\pi^{+}\pi^{-}\mu^{+}\mu^{-}$ decay, leading to an upper limit on the branching fraction of $\mathcal{B}(\eta\rightarrow\pi^{+}\pi^{-}\mu^{+}\mu^{-})<4.0\times10^{-7}$ at the 90\% confidence level.
Furthermore, the $CP$-violation asymmetry parameter is found to be $\mathcal{A}_{CP}(\eta\rightarrow\pi^{+}\pi^{-}e^{+}e^{-})=(-4.04\pm4.69_{\rm{stat.}}\pm0.14_{\rm{syst.}})\%$, showing no evidence of $CP$-violation with current statistics.
Additionally, we extract the transition form factor from the decay amplitude of $\eta\rightarrow\pi^+\pi^-e^+e^-$. 
Finally, axion-like particles are searched for via the decay $\eta\rightarrow\pi^+\pi^-a, a\rightarrow e^+e^-$, and upper limits on this branching fraction relative to that of $\eta\rightarrow\pi^+\pi^-e^+e^-$ are presented as a function of the axion-like particle mass in the range $5-200\ \mathrm{MeV}/c^{2}$.
\end{abstract}

\maketitle

\section{INTRODUCTION}
The decays $\eta\rightarrow\pi^{+}\pi^{-}l^+l^-$ ($l=e$ or $\mu$) involve contributions from the box anomaly of quantum chromodynamics (QCD) and have been studied using various models, such as the hidden gauge~\cite{chiral_unitary_approach}, the chiral unitary approach (Unitary $\chi$PT)~\cite{chiral_unitary_approach}, and the vector meson dominance (VMD)~\cite{1010_2378} models. 
Table~\ref{theorybr} displays the theoretical predictions and experimental results for the branching fractions ($\mathcal{B}$) of the $\eta\rightarrow\pi^{+}\pi^{-}l^+l^-$ decays.
Electromagnetic decays of mesons depend on the coupling of photons to the electric charge distribution of the quark fields.   
This electromagnetic meson structure is described by the transition form factor (TFF), which is a function of the momentum transfer.
More importantly, the TFF determines the size of certain hadronic quantum corrections in the anomalous magnetic moment of the muon, $(g-2)_\mu$~\cite{Danilkin:2019mhd}.
\begin{table}[htbp]
    \setlength{\abovecaptionskip}{0pt}%
    \setlength{\belowcaptionskip}{10pt}%
    \caption{Different theoretical predictions and previous experimental measurements of $\mathcal{B}(\eta\rightarrow\pi^{+}\pi^{-}l^{+}l^{-})$.}
    \label{theorybr}
    \centering
    \renewcommand\arraystretch{1.2}
    \scalebox{1.0}{
    \begin{tabular}{ccc}
        \hline\hline
        & \makecell[c]{$\mathcal{B}(\eta\rightarrow\pi^{+}\pi^{-}e^{+}e^{-})$\\ $(10^{-4})$} & \makecell[c]{$\mathcal{B}(\eta\rightarrow\pi^{+}\pi^{-}\mu^{+}\mu^{-})$\\  $(10^{-9})$} \\\hline
        Unitary $\chi$PT~\cite{chiral_unitary_approach} & $2.99^{+0.06}_{-0.09}$      & $7.50^{+1.80}_{-0.70}$ \\
        Hidden gauge~\cite{1010_2378}     & $3.14\pm0.17$              & $8.65\pm0.39$          \\
        VMD~\cite{1010_2378}              & $3.02\pm0.12$              & $8.64\pm0.25$          \\
        SM~\cite{Zillinger:2022eva}  & $2.65\pm0.17$    &$6.36\pm0.39$ \\
        CMD-2~\cite{CMD2_2001}    & $3.7^{+2.5}_{-1.8}\pm3.0$       & $...$   \\
        WASA~\cite{WASA_2008}    & $4.3^{+0.2}_{-1.6}\pm0.4$       & $<3.6\times10^{5}$                 \\
        KLOE~\cite{KLOE_2009}             & $2.68\pm0.09\pm0.07$ & $...$             \\
        \hline\hline
    \end{tabular}}
\end{table}

In the year 2000, the KTeV experiment observed a significant $CP$ asymmetry in the distribution of the T-odd angle $\varphi$ in $K^{0}_{L}\rightarrow\pi^{+}\pi^{-}e^{+}e^{-}$ decay~\cite{KL_pipiee_1}, where $\varphi$ is the angle between the $e^{+}e^{-}$ decay plane and the $\pi^{+}\pi^{-}$ decay plane in the $K^{0}_{L}$ center-of-mass system.
However, so far there is no experimental indication of $CP$-violation in flavor-conserving reactions.
The $\eta\rightarrow\pi^{+}\pi^{-}e^+e^-$ decay provides an opportunity to search for an analogous $CP$-violation asymmetry in the angle distribution between the $e^{+}e^{-}$ and $\pi^{+}\pi^{-}$ decay planes. 
This has been explored in $\eta\rightarrow\pi^{+}\pi^{-}e^+e^-$ decay in the KLOE experiment, and the $CP$-violation asymmetry is determined to be $\mathcal{A}_{CP}=(-0.6\pm2.5\pm1.8)\%$~\cite{KLOE_2009}, which is consistent with zero. 

The hadronic decay channels of the $\eta$ meson could also provide signals of the QCD axion, a dark photon or other axion-like particles (ALPs)~\cite{axion_etaTopipia_2020,Alves:2024dpa} that couple with hadrons.
Hints of a new light bosonic state with mass around $17\ \mathrm{MeV/c^2}$~\cite{axion_Be8_2016, axion_He4_2019} were observed by the ATOMKI Collaboration via measuring the angle between $e^+e^-$ pairs~\cite{BUTTAZZO2021136310, Ellwanger:2016wfe, Liu:2021wap},
and a light pseudoscalar particle $a$ decaying to $e^+e^-$ was proposed to explain the ATOMKI anomaly. 
The ALPs could also cause a deviation from the expected value of the electron anomalous magnetic moment~\cite{doi:10.1126/science.aap7706, Morel:2020dww, NA64:2021xzo}.

The ten billion $J/\psi$ events collected with the BESIII detector during 2009-2019~\cite{BESIII_2021_njpsi} offer an excellent opportunity to measure the TFF, and to search for $CP$-violation and hadronically coupled ALPs.  

\section{BESIII DETECTOR}
The BESIII detector~\cite{BESIII_2009_detector} records symmetric $e^+e^-$ collisions 
provided by the BEPCII storage ring~\cite{BESIII_2016_detector}
in the center-of-mass energy range from 2.0 to 4.95~GeV,
with a peak luminosity of $1 \times 10^{33}\;\text{cm}^{-2}\text{s}^{-1}$ 
achieved at $\sqrt{s} = 3.77\;\text{GeV}$. 
BESIII has collected large data samples in this energy region~\cite{BESIII:2020nme}. The cylindrical core of the BESIII detector covers 93\% of the full solid angle and consists of a helium-based multilayer drift chamber~(MDC), a plastic scintillator time-of-flight
system~(TOF), and a CsI(Tl) electromagnetic calorimeter~(EMC),
which are all enclosed in a superconducting solenoidal magnet
providing a 1.0~T magnetic field.
The magnetic field was 0.9~T in 2012, which affects 10\% of the total $J/\psi$ data.
The solenoid is supported by an
octagonal flux-return yoke with resistive plate counter muon
identification modules interleaved with steel. 
The charged-particle momentum resolution at $1~{\rm GeV}/c$ is
$0.5\%$, and the 
${\rm d}E/{\rm d}x$
resolution is $6\%$ for electrons
from Bhabha scattering. The EMC measures photon energies with a
resolution of $2.5\%$ ($5\%$) at $1$~GeV in the barrel (end cap)
region. The time resolution in the TOF barrel region is 68~ps, while
that in the end cap region was 110~ps. 
The end cap TOF
system was upgraded in 2015 using multigap resistive plate chamber
technology, providing a time resolution of
60~ps~\cite{BESIII_2019_detector}.
About 87\% of the data used here benefits from this upgrade.

\section{DATA SAMPLE AND MONTE CARLO SIMULATION}
This analysis is based on $(10087\pm44)\times10^{6}$ $J/\psi$ events collected with the BESIII detector from 2009 to 2019~\cite{BESIII_2021_njpsi}; it uses the radiative decay $J/\psi\rightarrow\gamma\eta$, resulting in a sample of about $1.1\times10^{7}$ $\eta$ events.

The estimation of backgrounds and the determination of detection efficiencies are performed with Monte Carlo (MC) simulations. The BESIII detectors are modeled with GEANT4~\cite{GEANT4_2002_1, GEANT4_2006_2,Huang:2022wuo}. The production of the $J/\psi$ resonance is implemented with the MC event generator KKMC~\cite{kkmc_1999_1, kkmc_2000_2}, while the decays are simulated by EVTGEN~\cite{EVTGEN_2008, Lange:2001uf}. The possible backgrounds are studied using a sample of $J/\psi$ inclusive events in which the known decays of $J/\psi$ are modeled with branching fractions set to the world average values in the PDG~\cite{ParticleDataGroup:2024cfk}, while the unknown decays are generated with the LUNDCHARM model~\cite{LUNDCHARM_2000}. 
For this analysis, specific generators for the $\eta\rightarrow\pi^{+}\pi^{-}l^+l^-$~\cite{Event_generators_2012_etap_pipill}, $\eta\rightarrow\pi^{+}\pi^{-}\pi^{0}$~\cite{generator_2018},
$\eta\rightarrow\gamma\pi^{+}\pi^{-}$~\cite{generator_2018}, $\eta\rightarrow\gamma e^{+}e^{-}$~\cite{generator_2018}, $\eta'\rightarrow\eta\pi^{+}\pi^{-}$~\cite{generator_2018} decays are developed based on theoretical amplitudes.

\section{Event selection and background analysis}

The final state of interest is studied through the decay chain $J/\psi\rightarrow\gamma\eta$, $\eta\rightarrow\pi^{+}\pi^{-}l^{+}l^{-}$. 
Each event is required to contain at least one good photon candidate, and four charged track candidates with a total charge of zero. 
Charged tracks detected in the MDC are required to be within a polar angle $(\theta)$ range of $|\cos \theta| \leq 0.93$, where $\theta$ is defined with respect to the $z$-axis, which is the symmetry axis of the MDC.
The distance of closest approach to the interaction point (IP) must be less than 10 cm along the $z$-axis and less than 1 cm in the transverse plane. 
Photon candidates are identified using showers in the EMC.
The deposited energy of each shower must be larger than 25 MeV in the barrel region ($|\cos \theta| < 0.8$) or larger than 50 MeV in the end cap region ($(0.86 < |\cos \theta|< 0.92)$).
To exclude showers that originate from charged tracks, the angle subtended by the EMC shower and the position of the closest charged track at the EMC must be greater than 15 degrees as measured from the IP.
To suppress electronic noise and showers unrelated to the event, the difference between the EMC time and the event start time is required to be within [0, 700] ns.

For each candidate, particle identification (PID) is performed using the TOF and d$E$/d$x$ information, and a four-constraint (4C) kinematic fit is executed imposing energy and momentum conservation under the hypothesis of $\gamma\pi^{+}\pi^{-}e^{+}e^{-}$ or $\gamma\pi^{+}\pi^{-}\mu^{+}\mu^{-}$ final state.
A summed chi-square, $\chi^2_{\rm{sum}}=\chi^2_{\rm{4C}}+\sum_{i=1}^4\chi^2_{{\rm PID}(i)}$, is calculated from the chi-square of the 4C kinematic fit ($\chi^2_{\rm{4C}}$) and PID ($\chi^2_{\rm{PID}}$). 
For each event, the hypothesis with the smallest $\chi^2_{\rm{sum}}(\pi^{+}\pi^{-}e^{+}e^{-})$ or $\chi^2_{\rm{sum}}(\pi^{+}\pi^{-}\mu^{+}\mu^{-})$ is kept for the further analysis. 

\begin{figure}[htbp]
\centering
\begin{subfigure}{}
\includegraphics[scale=0.42]{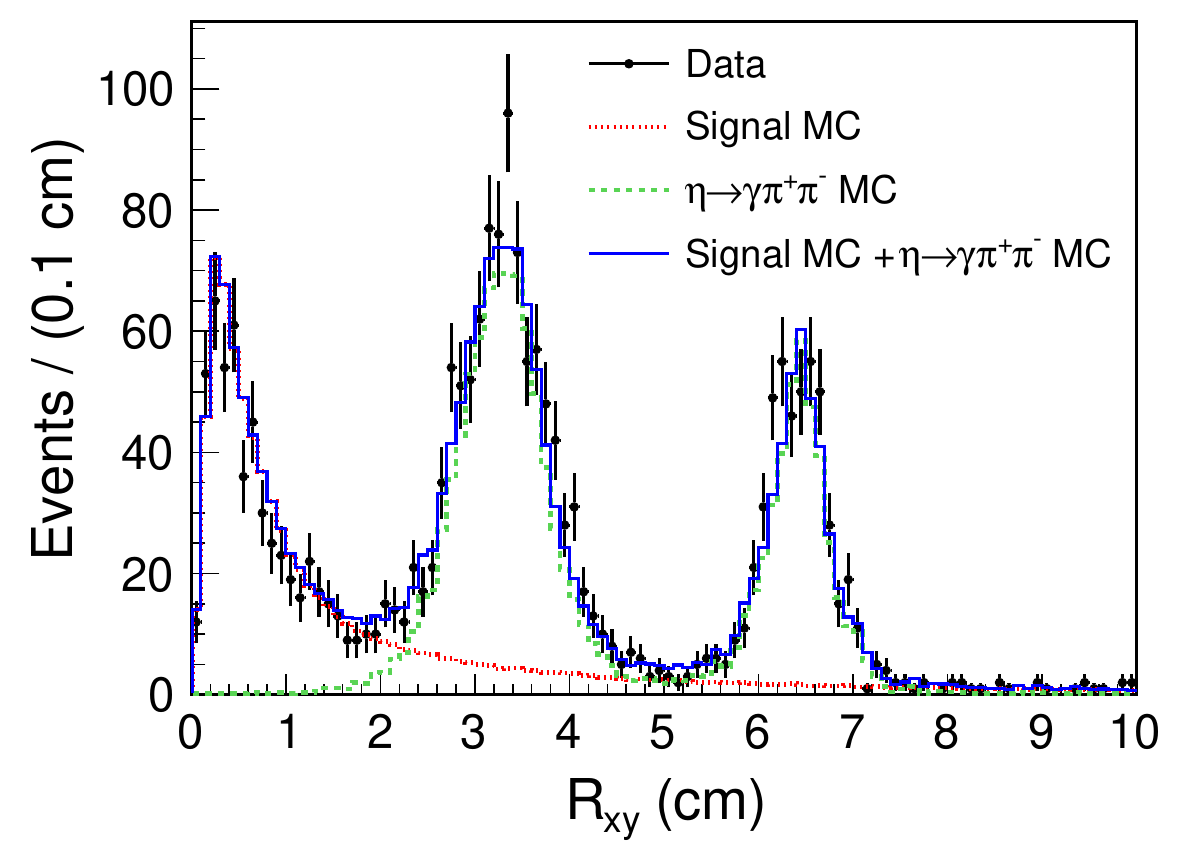}
\end{subfigure}
\caption{    
The distribution of $R_{xy}$.
The dots with error bars represent the data, the red dashed histogram is the MC signal shape, the green dashed histogram is the $J/\psi\rightarrow\gamma\eta, \eta\rightarrow\gamma\pi^+\pi^-$ MC shape, and the blue solid histogram is the sum of the MC signal and MC background from $J/\psi\rightarrow\gamma\eta, \eta\rightarrow\gamma\pi^+\pi^-$. 
}
\label{Mee_Rxy}
\end{figure}

For the decay $\eta\rightarrow\pi^{+}\pi^{-}e^{+}e^{-}$, a requirement of $\chi^{2}_{\rm{sum}}(\pi^+\pi^-e^+e^-)<50$ is imposed.
The primary peaking background comes from the $\eta\rightarrow\gamma\pi^+\pi^-$ events,
where the photon converts to an $e^+e^-$ pair.
To exclude these background events, a photon conversion finder algorithm~\cite{GamConv} is applied to the selected $e^+e^-$ pair.
The distribution of $R_{xy}$, defined as the projected distance from the $e^+e^-$-vertex position to the IP  in the $x-y$ plane~\cite{GamConv}, is shown in Fig.~\ref{Mee_Rxy}.

\begin{figure*}[htbp]
\centering
\begin{subfigure}{}
\includegraphics[scale=0.42]{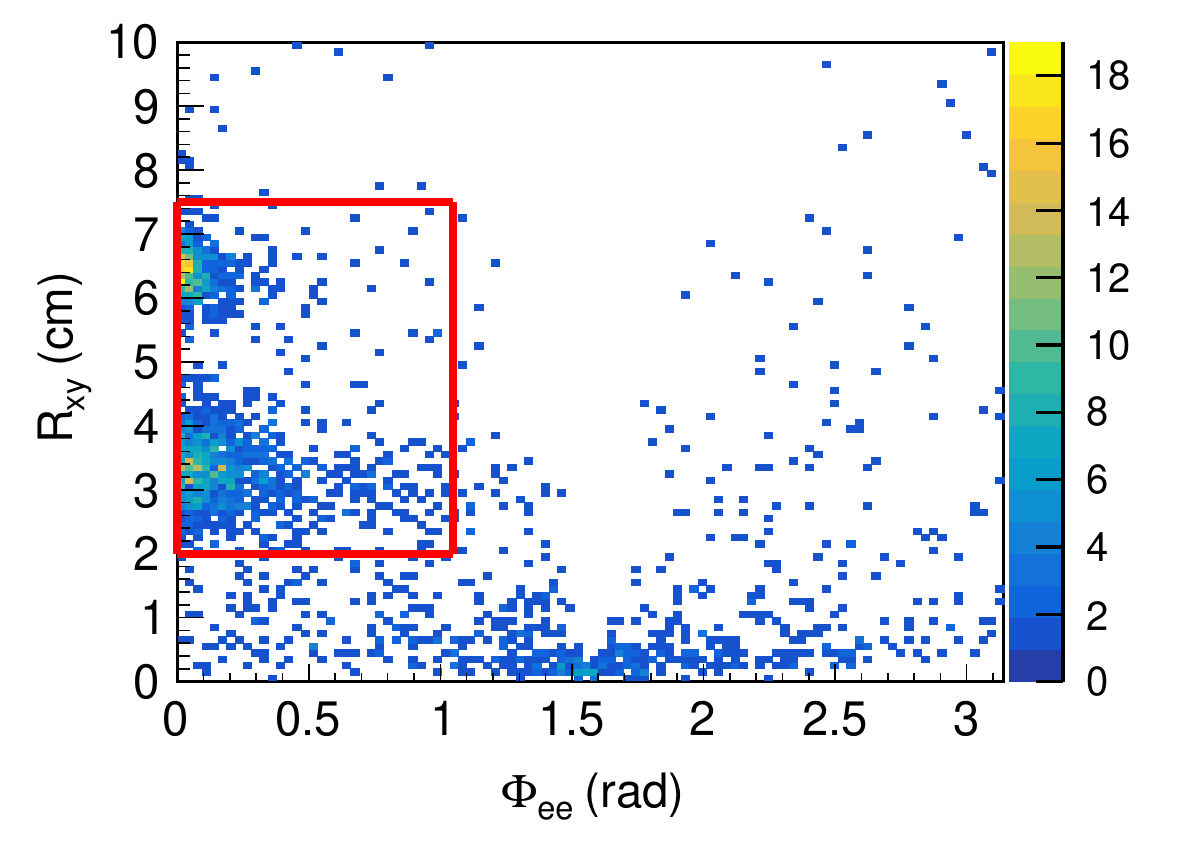}
\put(-120,140){(a)}
\end{subfigure}
\begin{subfigure}{}
\includegraphics[scale=0.42]{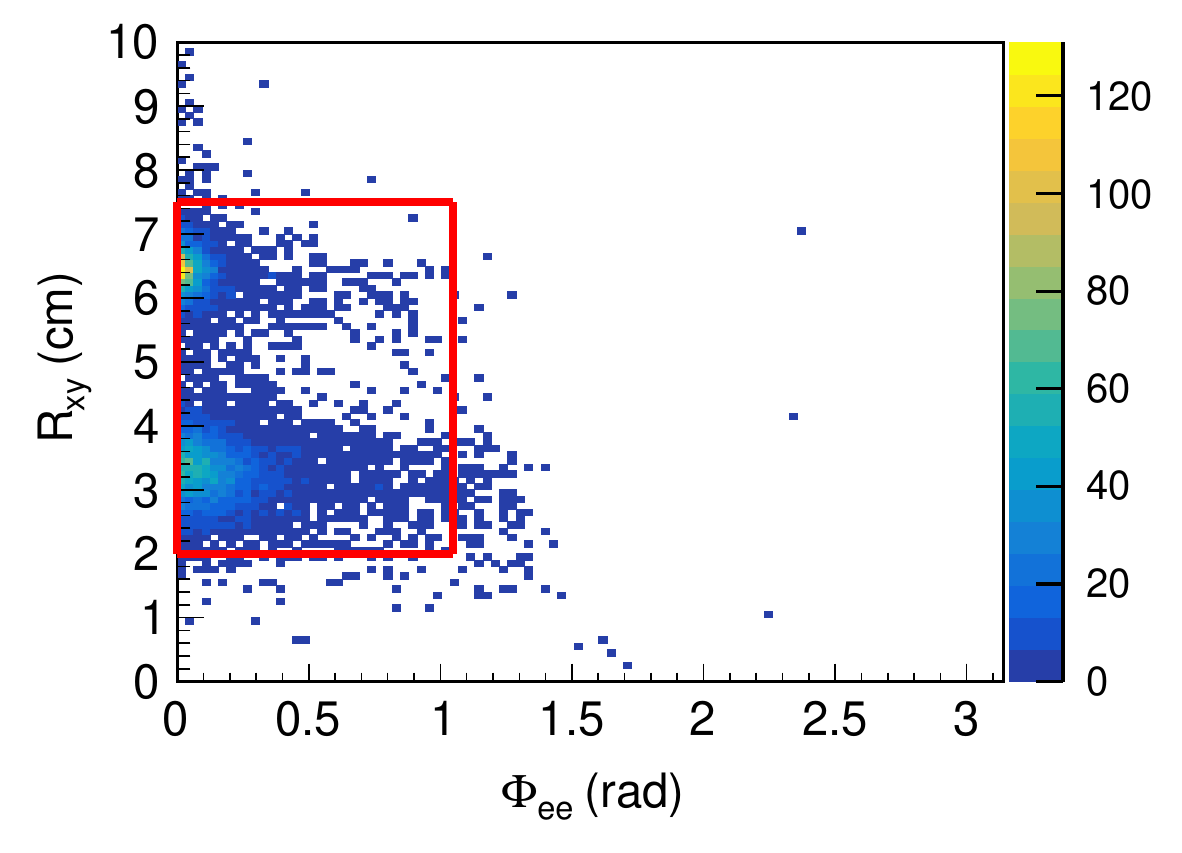}
\put(-120,140){(b)}
\end{subfigure}
\caption{The photon conversion veto criterion in the $\Phi_{ee}$ vs. $R_{xy}$ plane, shown with events from (a) data and (b) $\eta\rightarrow\gamma\pi^+\pi^-$ background MC sample. Events inside the red rectangle are rejected.}
\label{scap_gamma_conversion}
\end{figure*}

Photon conversion events are identified based on their proximity to the beam pipe ($R_{xy}\approx3.5$ cm) and the inner wall of the MDC ($R_{xy}\approx6.5$ cm).
The observable $\Phi_{ee}$ is the lab-frame opening angle of the $e^+e^-$ pair~\cite{PhysRevC.81.034911}, and the two-dimensional distribution of $\Phi_{ee}$ vs. $R_{xy}$ is shown in Fig.~\ref{scap_gamma_conversion}.
For photon conversion events, $\Phi_{ee}$ is expected to be close to zero.
After the application of the photon conversion veto $\Phi_{ee}<\frac{\pi}{3}$ when $2.0\ \mathrm{cm} <R_{xy}<7.5\ \mathrm{cm}$, most conversion events are rejected.
The remaining background events from the inclusive MC sample are studied, 
and the dominant background channels listed in Table~\ref{backgrounds}.
Dedicated exclusive MC samples are then generated to better estimate their contributions.

\begin{table}[htbp]
    \setlength{\abovecaptionskip}{0pt}%
    \setlength{\belowcaptionskip}{10pt}%
    \caption{Main background processes and normalized yields of the $\eta\rightarrow\pi^+\pi^-e^+e^-$ decay. For the $\mathcal{B}$ determination, ``$-$'' indicates that the number of background events are float in the fit. ``Other'' refers to the analyses for TFF, $\mathcal{A}_{CP}$, and an  ALP. A more detailed explanation is provided in the subsequent sections.}
    \label{backgrounds}
    \centering
    \footnotesize
    \renewcommand\arraystretch{1.2}
    \scalebox{1.0}{
    \begin {tabular}{ccc}  \hline\hline 
    \multirow{2}*{Background mode} & \multicolumn{2}{c}{Normalized yields} \\
    \cline{2-3} ~ & $\mathcal{B}$ & Other  \\ 
    \hline
    $J/\psi\rightarrow\gamma\eta', \eta'\rightarrow\pi^+\pi^-\eta, \eta\rightarrow\gamma e^+e^-$  &$85\pm6$    &$8\pm2$\\
    $J/\psi\rightarrow\gamma\eta, \eta\rightarrow\pi^+\pi^-\pi^0$                                 &$312\pm13$  &$18\pm2$\\
    $J/\psi\rightarrow\gamma\eta, \eta\rightarrow\gamma\pi^+\pi^-$                                &$42\pm3$    &$29\pm3$\\
    $ J/\psi\rightarrow\gamma\pi^+\pi^-\pi^+\pi^-$                                                &$-$   &$16\pm2$\\
    \hline\hline
    \end{tabular}}
\end{table}

\begin{figure}[htbp]
\centering
\subfigure{\includegraphics[scale=0.41]{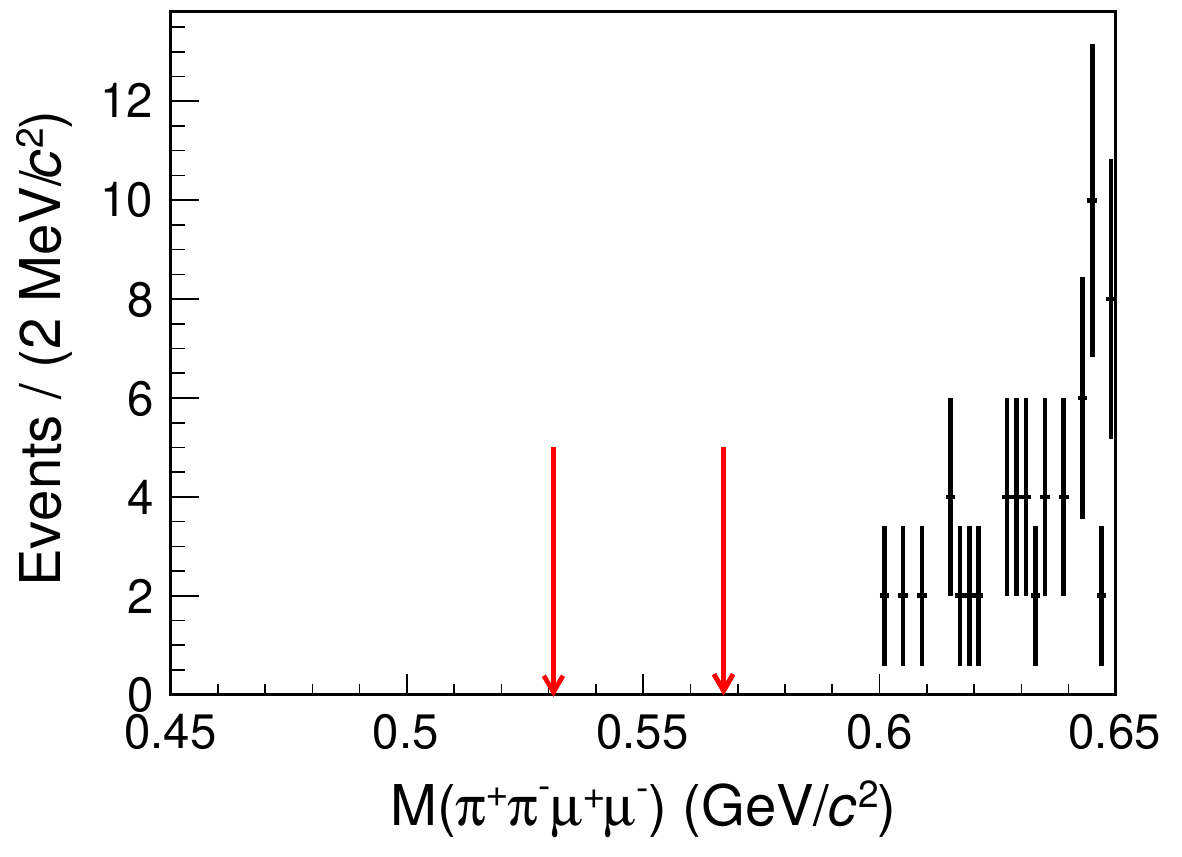}}
\caption{The $\pi^{+}\pi^{-}\mu^{+}\mu^{-}$ invariant mass distribution of selected candidates in data.}
\label{Mpipimumu}
\end{figure}

For the decay $\eta\rightarrow\pi^{+}\pi^{-}\mu^{+}\mu^{-}$, a requirement of $\chi^2_{\rm{sum}}(\pi^+\pi^-\mu^+\mu^-) < 50$ is imposed.
The signal window is determined by fitting the $\pi^+\pi^-\mu^+\mu^-$ invariant mass spectrum of the signal MC sample.
The signal shape is represented by a double Gaussian function, and the signal window is defined as $[\mu-3\sigma, \mu+3\sigma]$, where $\mu$ and $\sigma$ are the weighted mean and standard deviation from the fit. 
In this case, the signal window is set to $0.531\ \mathrm{GeV}/c^{2} < M(\pi^{+}\pi^{-}\mu^{+}\mu^{-}) < 0.567\ \mathrm{GeV}/c^{2}$, and there are no events in this $\eta$ mass region, as displayed in  Fig.~\ref{Mpipimumu}.

\section{Analysis of $\eta\rightarrow \pi^{+}\pi^{-}e^{+}e^{-}$}
\subsection{Branching Fraction Measurement}
\begin{figure}[htbp]
\centering
\subfigure{\includegraphics[scale=0.42]{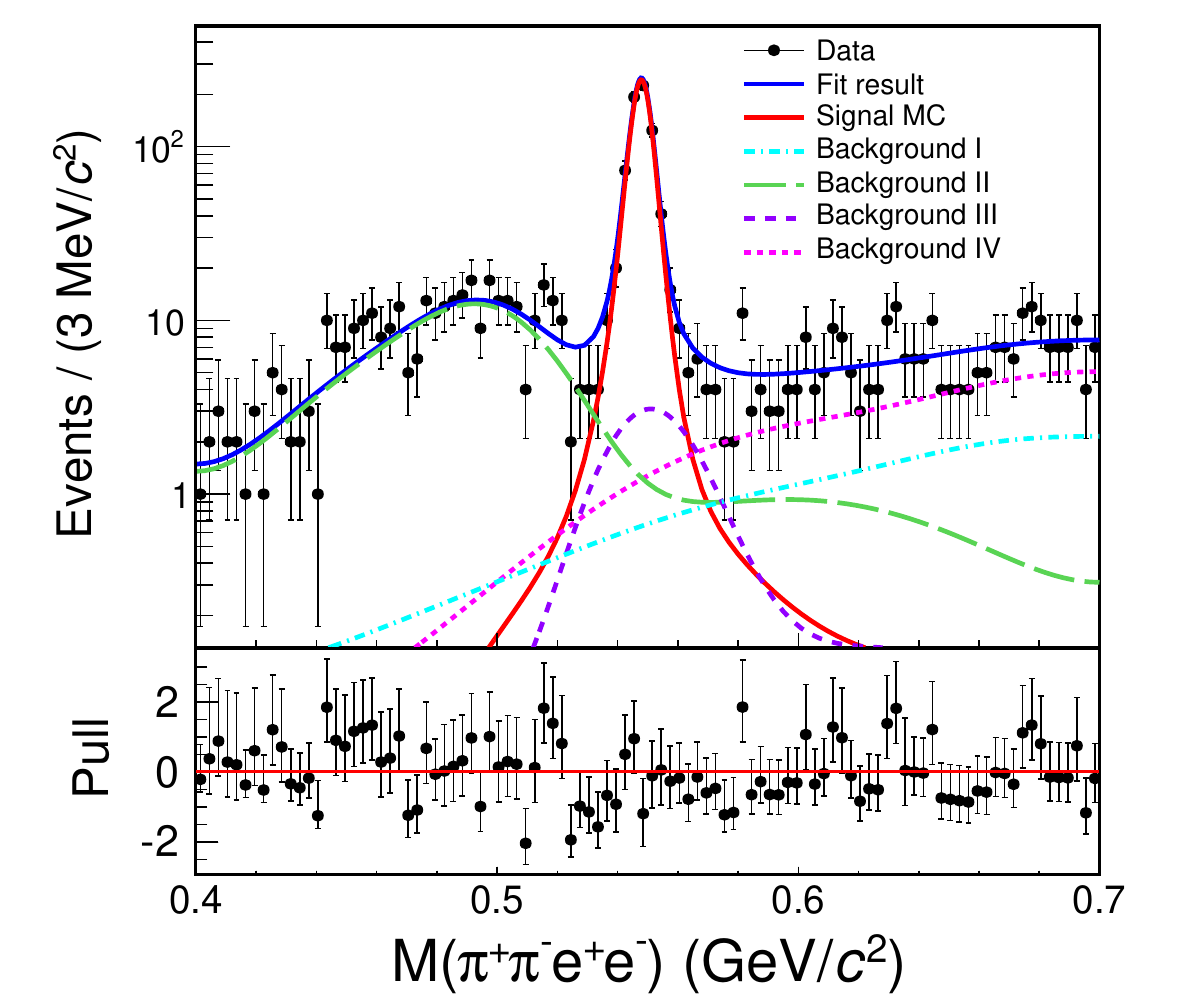}}
\caption{    
Fit to the invariant mass distribution of $\pi^+\pi^-e^+e^-$.
The dots with error bars represent the data, the red solid line is the MC signal shape, and the blue solid line is the total fit result. 
The blue dot-dash line (Background I) is the $J/\psi\rightarrow\gamma\eta',\eta'\rightarrow\pi^+\pi^-\eta,\eta\rightarrow\gamma e^+e^-$ MC shape.
The green long-dash line (Background II) is the $J/\psi\rightarrow\gamma\eta, \eta\rightarrow\pi^+\pi^-\pi^0$ MC shape.
The purple dashed line (Background III) is the $J/\psi\rightarrow\gamma\eta, \eta\rightarrow\gamma\pi^+\pi^-$ MC shape.
The pink short-dash line (Background IV) is the $J/\psi\rightarrow\gamma\pi^+\pi^-\pi^+\pi^-$ MC shape.
}
\label{fit_br}
\end{figure}

To extract the number of $\eta\rightarrow\pi^+\pi^-e^+e^-$ events, an unbinned maximum likelihood fit is performed to the $\pi^+\pi^-e^+e^-$ invariant mass spectrum. 
The signal and all the background shapes are taken from the MC simulations.
The number of background events from $J/\psi\rightarrow\gamma\pi^+\pi^-\pi^+\pi^-$ is determined from the fit, and all other background events are fixed according to the branching fractions from the PDG~\cite{ParticleDataGroup:2024cfk}, as summarized in Table~\ref{backgrounds}.
The fit yields $N_{\rm{sig}}=680\pm27$ signal events, and the result of the invariant mass of $\pi^+\pi^-e^+e^-$ is shown in Fig.~\ref{fit_br}.
The fit $\chi^{2}$ per degree of freedom ($ndf$) is $\chi^2/ndf=75.1/98$.  
The branching fraction is determined by
\begin{equation}
\begin{aligned}
\mathcal{B}(\eta\rightarrow\pi^+\pi^-e^+e^-)&=\frac{N_{\rm{sig}}}{N_{J/\psi} \, \mathcal{B}(J/\psi\rightarrow\gamma\eta) \, \varepsilon}.
\end{aligned}
\end{equation}
Here $N_{J/\psi}$ is the number of $J/\psi$ events~\cite{BESIII_2021_njpsi} and $\mathcal{B}(J/\psi\rightarrow\gamma\eta)$ is the branching fraction of $J/\psi\rightarrow\gamma\eta$~\cite{ParticleDataGroup:2024cfk}. 
With a detection efficiency of $\varepsilon=(20.12\pm0.04)\%$, the branching fraction of $\eta\rightarrow\pi^+\pi^-e^+e^-$ is calculated to be $(3.07\pm0.12) \times 10^{-4}$, where the uncertainty is statistical only.

\subsection{Form Factor Measurement}
\label{sec:TFF}
The form factor is written as  
\begin{equation}
\mathcal{M}(s_{\pi\pi}, s_{ee})=\mathcal{M}_{\eta} \times \mathrm{VMD}(s_{\pi\pi}, s_{ee}).
\end{equation}
Here $\mathcal{M}_{\eta}$ includes factors related to decay constants and pseudoscalar-octet mixing, and $\mathrm{VMD}(s_{\pi\pi}, s_{ee})$ is the VMD factor which included propagator effects.  
The four-momenta of the $\eta\rightarrow\pi^{+}\pi^{-}e^{+}e^{-}$ decay is  $P_{\eta}=P_{\pi^+}+P_{\pi^-}+P_{e^{+}}+P_{e^{-}}$, and the VMD function arguments are~\cite{KL_pipiee_2} 
\begin{equation}
\begin{aligned}
s_{\pi\pi}=(P_{\pi^+}+P_{\pi^-})^2, s_{ee}=(P_{e^{+}}+P_{e^{-}})^2.
\end{aligned}
\end{equation}
The VMD factor is derived from the VMD model with finite-width corrections~\cite{vmd_factor_1968_2} as
\begin{equation}
\begin{aligned}
    \mathrm{VMD}(s_{\pi\pi}, s_{ee}) = &1 - \frac{3}{4}(c_1-c_2+c_3) \\
    &+ \frac{3}{4}(c_1-c_2-c_3)\frac{m^2_V}{m^2_V-s_{ee}-im_V\Gamma(s_{ee})} \\
    &+ \frac{3}{2}c_3\frac{m^2_{V}}{m^2_{V}-s_{\pi\pi}-im^2_{V}\Gamma(s_{\pi\pi})} \\
    &\cdot\frac{m^2_V}{m^2_V-s_{ee}-im_V\Gamma(s_{ee})}.
\label{amp_vmd}
\end{aligned}
\end{equation}
Here $m_V$ is the mass of the vector meson and $\Gamma(s)$ is its total width~\cite{Event_generators_2012_etap_pipill} 
\begin{align}
    \Gamma(s) \,=\, \left(\frac{\Gamma_{\rho(770)} \, \sqrt{s}}{m_{V}}\right)\left(\frac{1-\frac{4m^2_{i}}{s}}{1-\frac{4m^2_{i}}{m^2_{V}}}\right)^{\frac{3}{2}} \Theta(s-4m^2_{i}) ,
\end{align}
where $i$ is $e\ \mathrm{or}\ \pi$, $\Gamma_{\rho(770)} = 149.1$ MeV is the width of $\rho(770)$, and $\Theta$ is
the Heaviside step function. 
By adjusting the values of the $c_i$-parameters~\cite{ci_parameters_2010},
one can switch between the various VMD models: 
(I) hidden gauge model ($c_1 - c_2 = c_3 = 1$); 
(II) full VMD model ($c_1 - c_2 = 1/3, c_3 = 1$); 

To extract the parameter $m_{V}$ in the VMD factor, an unbinned maximum likelihood fit to the $e^{+}e^{-}$ and $\pi^{+}\pi^{-}$ invariant mass spectra in data is performed using MINUIT~\cite{JAMES1975343}. 
In the fit, $M(\pi^{+}\pi^{-}e^{+}e^{-})$ is required to be in the $\eta$ mass region
of $(0.53-0.57) \mathrm{GeV}/c^{2}$.   
The backgrounds are generated by MC simulation with specific generators as described in Sec. III.
The number of background events of $J/\psi\rightarrow\gamma\pi^+\pi^-\pi^+\pi^-$ is obtained by fitting the $\pi^+\pi^-e^+e^-$ invariant mass spectrum, and other background yields, listed in Table~\ref{backgrounds}, are estimated using PDG branching fractions~\cite{ParticleDataGroup:2024cfk}. 
The probability of observing the $i^{th}$ event characterized by the measured four-momenta $\xi_i$ of final-state particles is
\begin{equation}
\begin{aligned}
    \mathcal{P}(\xi_i)=\frac{\vert\mathcal{A}(\xi_i)\vert^2 \, \varepsilon(\xi_i)}{\int\vert\mathcal{A}(\xi)\vert^2 \, \varepsilon(\xi) \, \mathrm{d}\xi},
\end{aligned}
\end{equation}
where $\mathcal{A}$ is the amplitude and $\varepsilon(\xi_i)$ is the detection efficiency.

The normalization factor in the denominator is computed by integrating over all kinematic variables.
The fit is done by minimizing the negative log-likelihood value
\begin{equation}
\begin{aligned}
  -\ln{\mathcal{L}} = -\omega' [ &
  \sum^{N_{\mathrm{data}}}_{i=1}\ln{\mathcal{P}(\xi_i)}-\omega_{\mathrm{bkg1}}\sum^{N_{\mathrm{bkg1}}}_{j=1}\ln{\mathcal{P}(\xi_j)}\\
  & -\omega_{\mathrm{bkg2}}\sum^{N_{\mathrm{bkg2}}}_{k=1}\ln{\mathcal{P}(\xi_k)}-...\, ],
\end{aligned}
\end{equation}
where $i$ runs over all accepted events, and $j$, $k$, ... run over the other considered background events.
Their corresponding numbers of events are denoted by $N_{\mathrm{data}}$, $N_{\mathrm{bkg1}}$ and $N_{\mathrm{bkg2}}$.
$\omega_{\mathrm{bkg1}}=\frac{N'_{\mathrm{bkg1}}}{N_{\mathrm{bkg1}}}$ and $\omega_{\mathrm{bkg2}}=\frac{N'_{\mathrm{bkg2}}}{N_{\mathrm{bkg2}}}$ are the weights of the backgrounds, where $N'_{\mathrm{bkg1}}$ and $N'_{\mathrm{bkg2}}$ are the contributions according to individual branching fractions taken from the PDG.
To obtain an unbiased uncertainty estimation, the normalization factor $\omega'$~\cite{Langenbruch:2019nwe} used is 
\begin{equation}
\omega'=\frac{N_{\mathrm{data}}-N_{\mathrm{bkg1}}\omega_{\mathrm{bkg1}}-N_{\mathrm{bkg2}}\omega_{\mathrm{bkg2}}+...}{N_{\mathrm{data}}+N_{\mathrm{bkg1}}\omega^{2}_{\mathrm{bkg1}}+N_{\mathrm{bkg2}}\omega^{2}_{\mathrm{bkg2}}+...}.
\end{equation}

\begin{figure*}[htbp]
\centering
\begin{subfigure}{}
\includegraphics[scale=0.41]{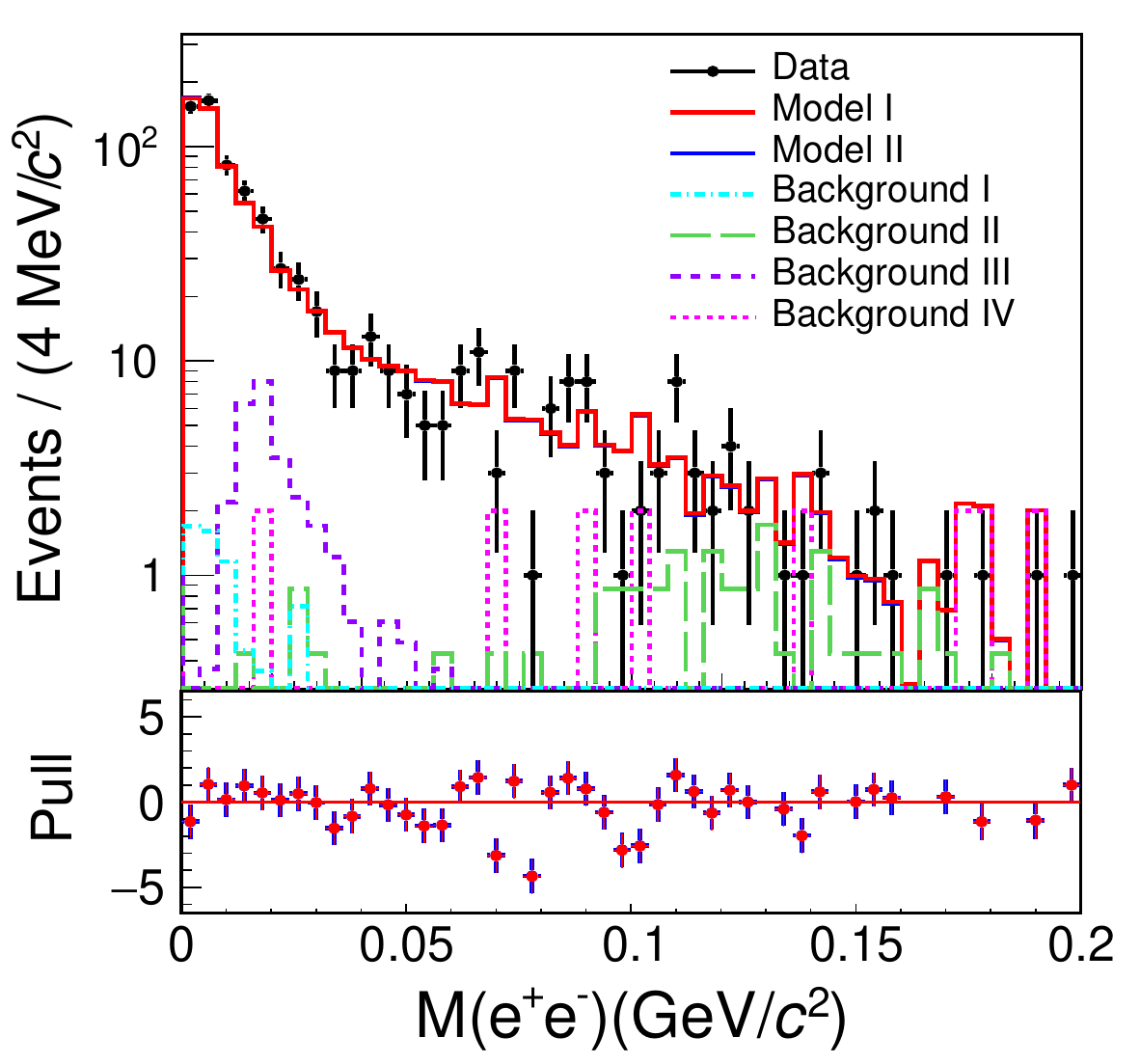}
\put(-60,140){(a)}
\end{subfigure}
\begin{subfigure}{}
\includegraphics[scale=0.41]{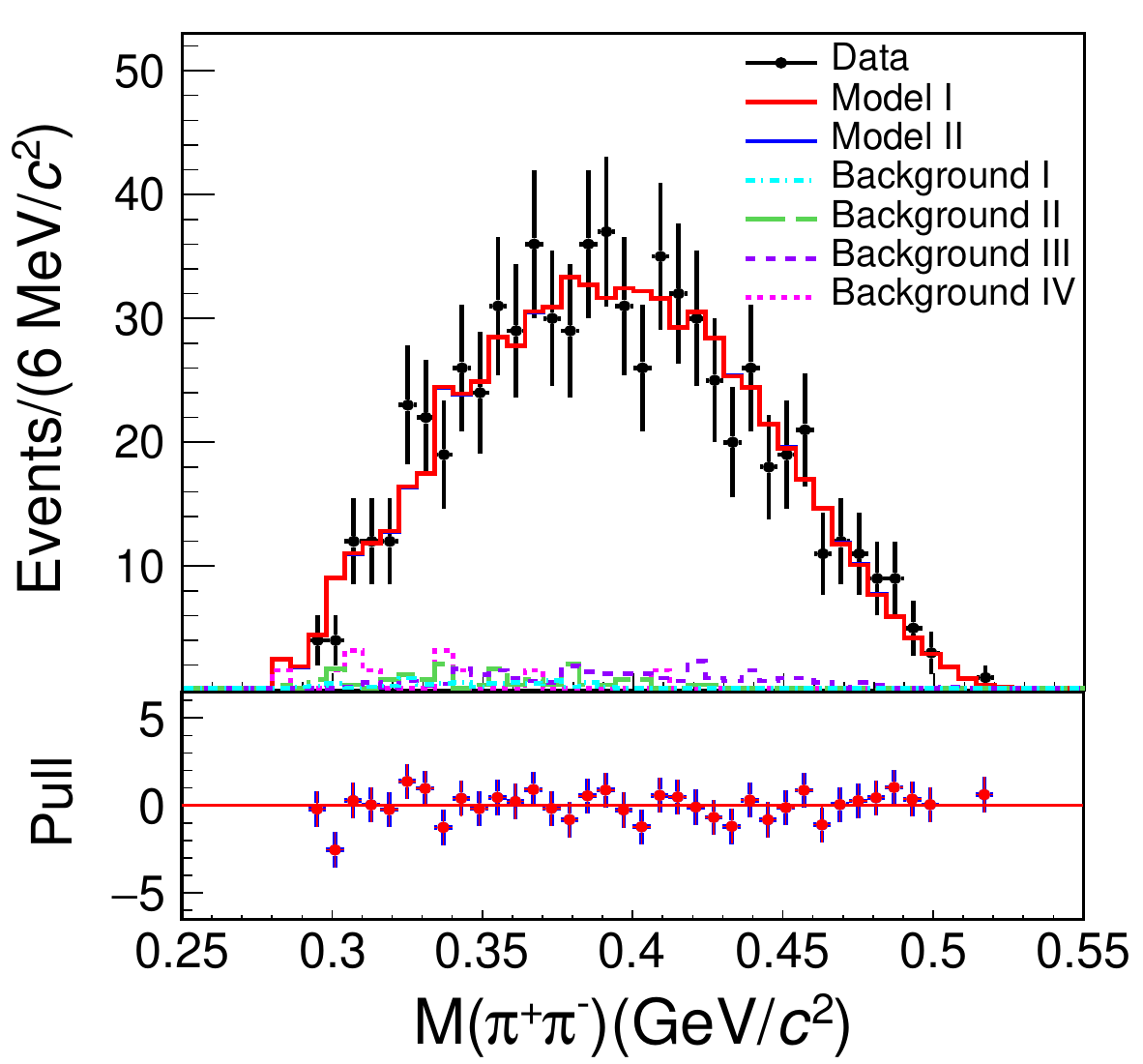}
\put(-60,140){(b)}
\end{subfigure}
\caption{Fit to the invariant mass distribution of the (a) $e^+e^-$ and (b) $\pi^+\pi^-$ pairs for the $\eta\rightarrow\pi^+\pi^-e^+e^-$ decay mode. 
The points with error bars represent the data. The red and blue solid histogram is the total fit result. 
The light blue dashed histogram (Background I) is the $J/\psi\rightarrow\gamma\eta',\eta'\rightarrow\pi^+\pi^-\eta,\eta\rightarrow\gamma e^+e^-$ MC shape.
The green dashed histogram (Background II) is the $J/\psi\rightarrow\gamma\eta, \eta\rightarrow\pi^+\pi^-\pi^0$ MC shape.
The purple dashed histogram (Background III) is the $J/\psi\rightarrow\gamma\eta, \eta\rightarrow\gamma\pi^+\pi^-$ MC shape.
The pink dotted histogram (Background IV) is the $J/\psi\rightarrow\gamma\pi^+\pi^-\pi^+\pi^-$ MC shape.
}
\label{pipiee_tff_11}
\end{figure*}

The fit results of Model I and Model II are both in good agreement with data; Fig.~\ref{pipiee_tff_11} shows the $\pi^+\pi^-$ and $e^+e^-$ invariant mass spectrum fits.
The results for $m_{V}$ in Model I and Model II are $m_{V}=749\pm54\ \mathrm{MeV}/c^{2}$ and $m_{V}=748\pm53\ \mathrm{MeV}/c^{2}$, respectively. 
The $\chi^2/ndf(\pi^+\pi^-,e^+e^-)$ are calculated to be (I) $73.6/41, 22.6/36$ and (II) $72.9/41, 22.6/36$.
A larger $\eta$ data sample, potentially obtainable from experiments like the Super $\tau-$Charm Facility~\cite{2023stcf}, is necessary to achieve a more precise measurement.

\subsection{$CP$-violation Asymmetry}
The fit function for the angular distribution, based on the squared decay amplitude~\cite{1010_2378}, is 
\begin{equation}
\begin{aligned}
    F(\varphi) = 1+a\cdot \sin^{2}{\varphi}+b\cdot \sin{2\varphi},
\end{aligned}
\end{equation}
where $1 + a \cdot \sin^{2}{\varphi}$ is a dominant contribution from the magnetic term, and $b \cdot \sin{2\varphi}$ is from the $CP$-violating interference term. Here, $\varphi$ is the angle between the decay planes of the $\pi^+\pi^-$ and $e^{+}e^{-}$ systems.
Then the asymmetry parameter $\mathcal{A}_{CP}$ is defined as

\begin{equation}
\begin{aligned}
  \mathcal{A}_{CP} &
  = \frac{1}{\Gamma} \int^{2\pi}_{0} \frac{d\Gamma}{d\varphi} \,
    \rm{sign}(\sin{2\varphi}) \, d\varphi \\
    & = \frac{\int^{\pi}_{-\pi} F(\varphi) \, \rm{sign}(\sin{2\varphi}) \,
      d\varphi}{\int^{\pi}_{-\pi}F(\varphi) \, d\varphi} \\
    & =  \frac{4b}{(2+a)\pi},
\label{Aphi}
\end{aligned}
\end{equation}
where $\mathrm{sign}(x)$ is the sign function, $\mathrm{sign}(x) = x/|x|$.

\begin{figure}[htbp]
\centering
\subfigure{\includegraphics[scale=0.41]{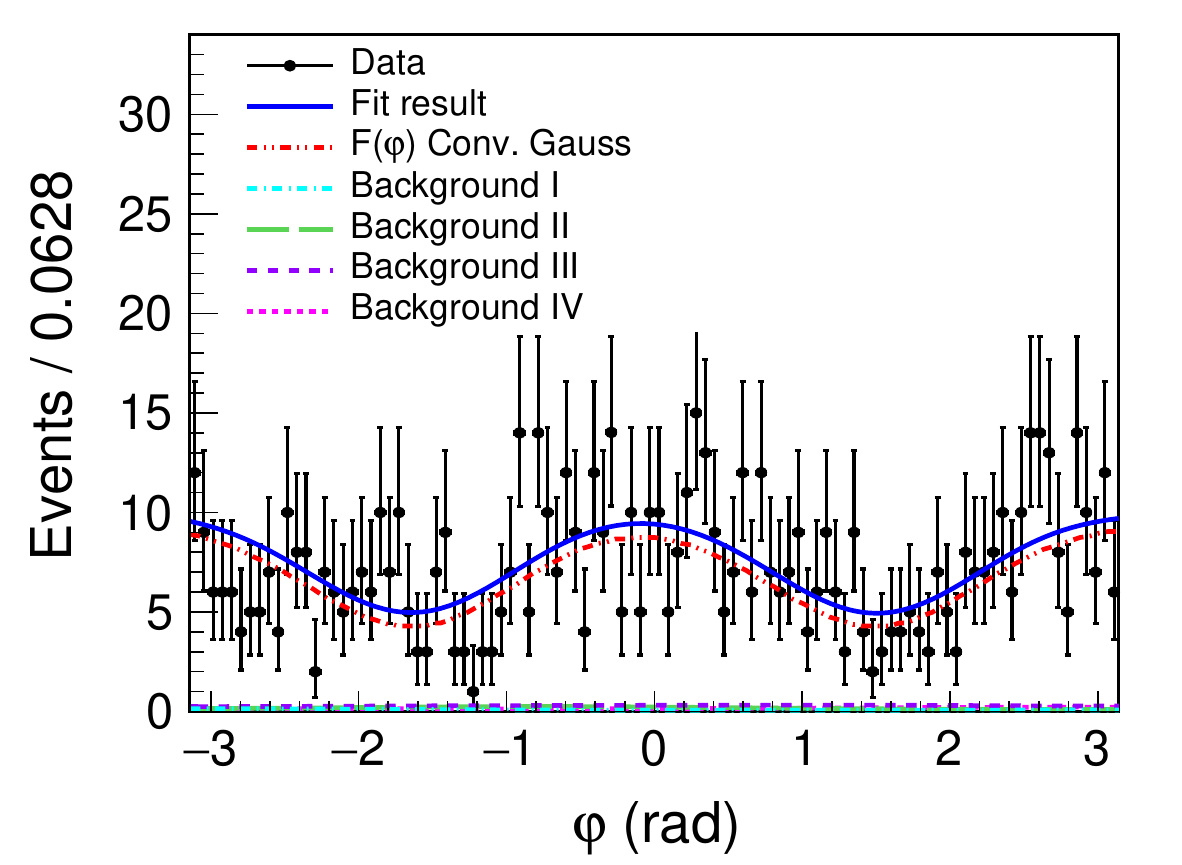}}
\caption{    
Fit to the distribution of plane angle $\varphi$ for $\eta\rightarrow\pi^+\pi^-e^+e^-$ decay mode.
The points with error bars represent the data, the blue solid histogram is the total fit result, and the red dashed line represents the function shape of $F(\varphi)$ convolved with a Gaussian function. 
The light blue dashed histogram (Background I) is the $J/\psi\rightarrow\gamma\eta',\eta'\rightarrow\pi^+\pi^-\eta,\eta\rightarrow\gamma e^+e^-$ MC shape.
The green dashed histogram (Background II) is the $J/\psi\rightarrow\gamma\eta, \eta\rightarrow\pi^+\pi^-\pi^0$ MC shape.
The purple dashed histogram (Background III) is the $J/\psi\rightarrow\gamma\eta, \eta\rightarrow\gamma\pi^+\pi^-$ MC shape.
The pink dotted histogram (Background IV) is the $J/\psi\rightarrow\gamma\pi^+\pi^-\pi^+\pi^-$ MC shape.
}
\label{acp_fit}
\end{figure}

The background treatment follows the same method as the TFF measurement (Section~\ref{sec:TFF}), and the background yields are listed in Table~\ref{backgrounds}.
In the fit, the efficiency-corrected $F(\varphi)$ function is convolved with a Gaussian function to account for the $\varphi$ resolution.
The fit result of $\varphi$ is shown in Fig.~\ref{acp_fit}; 
the $\chi^2/ndf$ is $79.8/97$.
With the fitted parameters $a=-0.696\pm0.061$ and $b=-0.041\pm0.048$, $\mathcal{A}_{CP}$ is calculated to be
\begin{equation}
\begin{aligned}
    \mathcal{A}_{CP}(\eta\rightarrow\pi^+\pi^-e^+e^-)=(-4.04\pm4.69)\%,
\end{aligned}
\end{equation}
where the uncertainty is statistical only.

\section{Systematic uncertainties}
Sources of systematic uncertainty are summarized in Table~\ref{sumsys}, and their corresponding contributions are discussed in detail below.

\begin{itemize}
\item Number of $J/\psi$ events: The number of $J/\psi$ events is determined to be $(10087\pm44)\times10^6$ from counting the inclusive hadronic events; the uncertainty is 0.4\%~\cite{BESIII_2021_njpsi}.
\item Branching fraction of $J/\psi\rightarrow\gamma\eta$: The uncertainty of 1.2\% on the branching fraction of $J/\psi\rightarrow\gamma\eta$ is taken from the PDG~\cite{ParticleDataGroup:2024cfk}. 
\item MC statistics: The systematic uncertainty related to the finite statistics of the MC simulation used to obtain the overall reconstruction efficiency is calculated as $\sqrt{\frac{\varepsilon(1-\varepsilon)}{N}}$, where $\varepsilon$ is the detection efficiency and $N$ is the number of generated MC events of the signal process. The corresponding systematic uncertainty is 0.2\%.
\item MDC tracking: The data-MC efficiency difference for pion track-finding is studied using a control sample of $J/\psi\rightarrow\pi^+\pi^-\pi^0$.  
Since there is no specific decay process available for us to study the tracking of muons in the low momentum region and the muon and pion masses are similar, we use the pion 
study results for muons as well.  
For the MDC tracking efficiency of electrons, a mixed sample of $e^+e^-\rightarrow e^+e^-\gamma$ at the $J/\psi$ meson mass and $J/\psi\rightarrow e^+e^-\gamma_{FSR}$ is used, where FSR stands for final-state radiation. 
The data-MC difference, $\Delta_{\rm{syst.}}$, is extracted as a function of the particle momentum and the cosine of the polar angle. 
Subsequently, each event in the MC samples is re-weighted by a factor $(1+\Delta_{\rm{syst.}})$. 
The branching fraction and asymmetry parameter are recalculated with efficiencies determined from the re-weighted MC sample.
For the TFF measurement, a re-weighted MC sample is used to calculate the MC integral, and a new set of fit results is obtained by using the fit method as outlined in Section~\ref{sec:TFF}.
The differences from the original results are taken as the systematic uncertainties.

\item Photon detection efficiency: 
The photon detection efficiency in the EMC is studied using a control sample of $e^+e^-\rightarrow\gamma_{\mathrm{ISR}}\mu^+\mu^-$, where ISR stands for initial-state radiation.
The systematic uncertainty is taken as 0.5\%, which is the maximum difference in efficiency between data and MC simulation in both the barrel and end-cap regions.

\item PID: 
The pion PID efficiency for data agrees within 1.0\% with that of the MC simulation in the pion momentum region, as reported in~\cite{BESIII_2020_etap_pipimumu}. 
There is no specific decay process available for us to study the PID of muons in the low momentum region.
Because the muon mass is similar to the pion mass, 1.0\% is taken as the systematic uncertainty for the muon~\cite{BESIII_2020_etap_pipimumu}. Thus, 4.0\% is taken as the systematic uncertainty from PID for the  $\eta\rightarrow\pi^{+}\pi^{-}\mu^{+}\mu^{-}$ decay.

\item Kinematic fit: To investigate the systematic uncertainty associated with the kinematic fit, the track helix parameter correction method~\cite{BESIII_2013_kmfitfit} is used. 
Half of the difference in the detection efficiencies with and without the helix corrections is taken as the systematic uncertainty.

\item Combined PID and kinematic fit: A clean control sample of $J/\psi\rightarrow\pi^+\pi^-\pi^0,\pi^0\rightarrow\gamma e^+e^-$ is used to study the systematic uncertainty due to the requirement of $\chi^2_{\rm{sum}}(\pi^+\pi^-e^+e^-)<50$. 
This sample included both $\pi^{0}$ Dalitz decay and $\pi^0\rightarrow\gamma\gamma$ decay with one photon externally converting to an electron-positron pair.
Using the same approach as that used for the tracking efficiency, we perform a two-dimensional correction to the selection efficiency of $\chi^2_{\rm{sum}}(\pi^+\pi^-e^+e^-)<50$ as a function of the momentum of the electron and positron. The differences from the original results are taken as the systematic uncertainties.

\item Photon conversion rejection:
The control sample of $J/\psi\rightarrow\pi^+\pi^-\pi^0,\pi^0\rightarrow\gamma e^+e^-$ is used to evaluate the systematic uncertainty from the rejection of photon conversions. 
Using the same approach as that used for the tracking efficiency, we perform a two-dimensional correction to the photon conversion rejection efficiency as a function of the momenta of the electrons and positrons. The differences from the original results are taken as the systematic uncertainties.

\item Signal shape: We use a double Gaussian function instead of the signal MC shape to fit $\pi^{+}\pi^{-}e^{+}e^{-}$ mass spectrum.
The fit yields $658\pm27$ signal events and the systematic uncertainty is determined as the difference from the nominal fit to be $3.2\%$.

\item Generator model: A MC generator based on a theoretical calculation is used to determine the detection efficiency for $\pi^{+}\pi^{-}e^{+}e^{-}$ decays. 
The dependence of the detection efficiency on the form factor is evaluated by replacing the nominal $m_V = 775\ \mathrm{MeV}/c^{2}$ with the $m_V = 749\ \mathrm{MeV}/c^{2}$ measurement as described in Section~\ref{sec:TFF}.
The difference of the detection efficiency between original model and model (I) is taken as the uncertainty.

\item Background estimation: The background simulations are based on the amplitude analysis and are relatively precise.
Therefore, the systematic uncertainty from background models is neglected.
In the fit, the numbers of events for specific backgrounds are fixed according to the branching fractions from the PDG.
To estimate the effect of the uncertainties of the used branching fractions, a set of random numbers is generated within the uncertainty range of each branching fraction. Using these random scaling parameters, a series of fits to the invariant mass distributions of $\pi^{+}\pi^{-}e^{+}e^{-}$  are performed. The variance of the determined number of signal events and asymmetry parameters is taken as the systematic uncertainties.
In the TFF measurement, we perform alternative fits by changing the branching fraction by $-1.0\sigma$ and $+1.0\sigma$, and the largest difference from the nominal results is taken as the systematic uncertainty.

\item Resolution: To estimate the uncertainty from the resolution, we perform alternative fits by changing the resolution from $-1.0\sigma$ to $+1.0\sigma$ for the $\eta\rightarrow\pi^{+}\pi^{-}e^{+}e^{-}$ decay.
The largest difference from the nominal result is taken as the systematic uncertainty on the asymmetry parameter.

\item Width: In the TFF measurement, the width of $\rho(770)$ is taken as a constant.
The difference of the fit results between the cases of fixing and allowing the width to float is taken as the systematic uncertainty.

\end{itemize}

\begin{table}[ht]
    \setlength{\abovecaptionskip}{0pt}%
    \setlength{\belowcaptionskip}{10pt}%
    \caption{The systematic uncertainties for the $\eta\rightarrow\pi^+\pi^-l^+l^-$ (with $l=e$ or $\mu$) decays.}
    \label{sumsys}
    \centering
    \renewcommand\arraystretch{1.1}
    \scalebox{1.0}{
    \begin{tabular}{ccccc}
        \hline\hline
        $\multirow{2}*{Source}$ & \multicolumn{3}{c}{$l=e$ (\%)} & $l=\mu$ (\%) \\
        \cline{2-5} ~ & $\mathcal{B}$ & $\mathcal{A}_{CP}$ & $m_{V}$ & $\mathcal{B}$ \\ 
        \hline
        Number of $J/\psi$ events                     & $0.4$   & $...$     & $...$    & $0.4$ \\
        $\mathcal{B}(J/\psi\rightarrow\gamma\eta)$    & $1.2$   & $...$     & $...$    & $1.2$ \\
        MC statistics                                 & $0.2$   & $...$     & $...$    & $0.2$\\
        MDC tracking                                  & $4.1$   & $...$     & $1.0$    & $0.7$ \\
        Photon detection                              & $0.5$   & $...$     & $...$    & $0.5$ \\
        PID                                           & $...$   & $...$     & $...$    & $4.0$ \\
        Kinematic fit                              & $...$   & $...$     & $...$    & $0.5$ \\
        Combine PID and kinematic fit                 & $3.1$   & $0.9$     & $0.2$    & $...$\\
        Photon conversion rejection                        & $1.1$   & $1.6$     & $1.6$    & $...$ \\
        Signal shape                                  & $3.2$   & $...$     & $...$    & $...$ \\
        Generator model                               & $0.1$   & $...$     & $...$    & $...$ \\
        Background estimation                              & $0.5$   & $2.2$     & $0.0$    & $...$ \\
        Resolution                                    & $...$   & $1.8$     & $...$    & $...$ \\
        Width                                         & $...$   & $...$     & $0.3$    & $...$\\
        \hline
        Total                                         & $6.3$   & $3.4$     & $1.9$    & $4.3$ \\
        \hline\hline
    \end{tabular}}
\end{table}

\section{Upper limit of $\eta\rightarrow\pi^+\pi^-\mu^+\mu^-$}
No events are found in the $\eta$ mass region ($N_{\rm{sig}}=0$). 
The number of remaining background events from all $J/\psi$ events is normalized to be $N_{\rm{bkg}} = 0$. 
The detection efficiency is $\varepsilon_{\rm{sig}} = (45.65\pm0.08)\%$ and the total systematic uncertainty is $\Delta_{\rm{sys.}} = 4.3\%$. 
The TROLKE method is useful in situations with low statistical significance, and the limit calculations make use of the profile likelihood method~\cite{TRolke}.
After inserting all the above numbers in the TROLKE program, the upper limit of signal events is obtained to be $N^{\mathrm{UP}} = 4.4$ at the 90\% confidence level (C.L.).
Thus, the upper limit on the branching fraction at the 90\% C.L. is determined as
\begin{equation}
\begin{aligned}
    \mathcal{B} < \frac{N^{\mathrm{UP}}}{N_{J/\psi}\cdot\mathcal{B}(J/\psi\rightarrow\gamma\eta)} = 4.0 \times 10^{-7}.
\end{aligned}
\end{equation}

\section{Upper limits for ALPs in $\eta\rightarrow\pi^+\pi^-a, a\rightarrow e^+e^-$}
To study the unknown ALPs decaying into $e^{+}e^{-}$, we perform 40 simulations under the hypothesis $a\rightarrow e^{+}e^{-}$ by varying masses of $a$ in steps of $5$ MeV/$c^{2}$ from 0 to 200 MeV/$c^{2}$. 

The $a$ signal shape is obtained through MC simulation with an assumption of a negligible width.
The treatment of the background is the same as that of the TFF measurement in Section~\ref{sec:TFF}, and the normalized background yields are listed in Table~\ref{backgrounds}.
Then 40 unbinned maximum likelihood fits to the $e^{+}e^{-}$ invariant mass spectrum are performed.

We consider possible sources for multiplicative systematic uncertainties of the upper limits for ALPs, such as number of $J/\psi$ events (0.4\%), branching fraction of $J/\psi\rightarrow\gamma\eta$ (1.2\%), MC statistics (0.1\%), MDC tracking ($2.6\%-5.0\%$), photon detection efficiency (0.5\%), combined PID and kinematic fit (3.1\%) and photon conversion veto (1.1\%). 
The total multiplicative systematic uncertainties for each hypothesis of $a$ mass are $(4.4 - 6.1)\%$.
The additive systematic uncertainties are considered by alternative fit ranges and Background estimation. The maximum signal yield among the different fit scenarios is adopted as its upper limit.

\begin{figure}[htbp]
\centering
\subfigure{\includegraphics[scale=0.41]{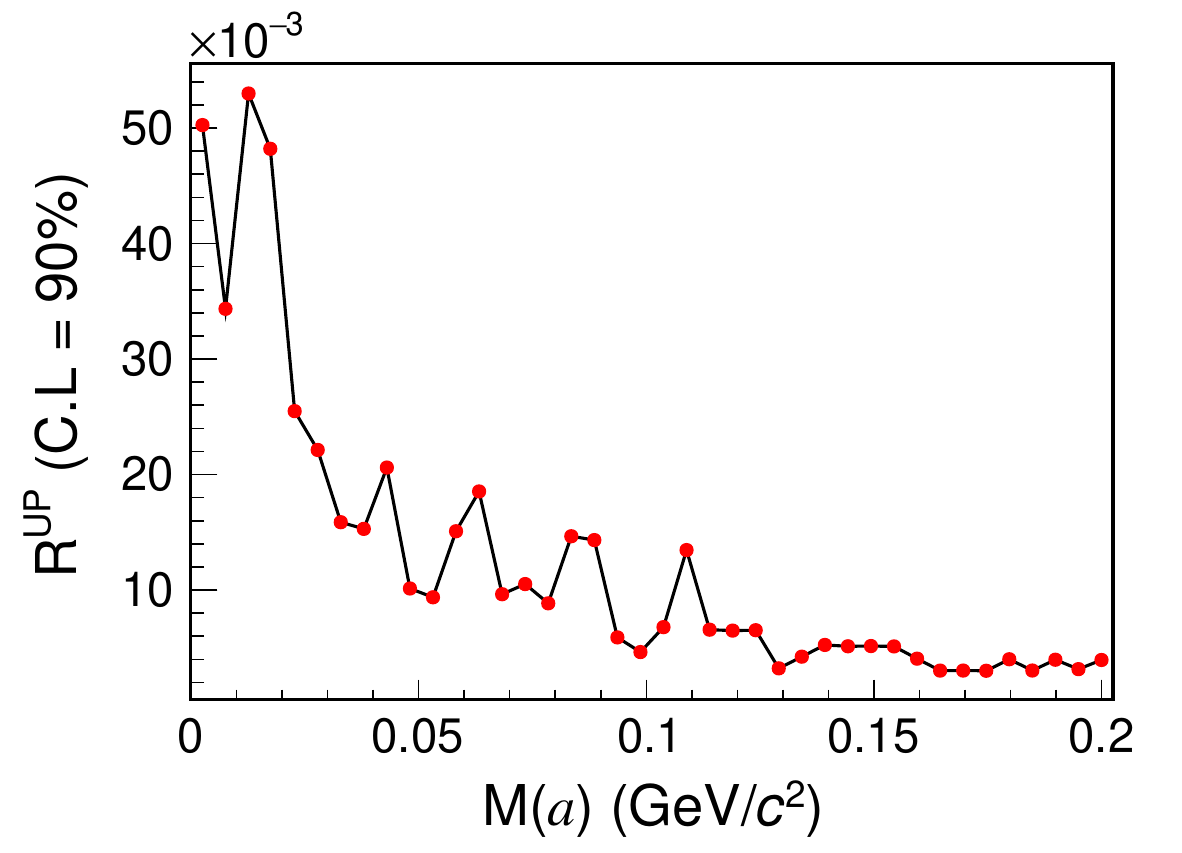}}
\caption{Upper limits on the relative width at the 90\% C.L. for different $a$ masses.}
\label{axion_fit}
\end{figure}

Since no evident $a$ signals are seen in the $M(e^+e^-)$ distribution, we compute the upper limits on the relative branching fraction, $R^{\mathrm{UP}}=\frac{\mathcal{B}(\eta\rightarrow\pi^{+}\pi^{-}a)\cdot\mathcal{B}(a\rightarrow e^{+}e^{-})}{\mathcal{B}(\eta\rightarrow\pi^{+}\pi^{-}e^{+}e^{-})}$, at the 90\% C.L. as a function of $M(a)$.
The upper limits on the number of $a$ signal events at the 90\% C.L. are obtained according to the Bayesian method~\cite{Zhu:2008ca} by smearing the likelihood curve using a Gaussian function with a width of the systematic uncertainty as
\begin{equation}
L'(N) = \int^{1}_{0} L\left(\frac{S}{\hat{S}}N\right)\mathrm{exp}\left[-\frac{(S-\hat{S})^{2}}{2\sigma_{S}^{2}}\right]\mathrm{d}S,
\end{equation}
where $L$ and $L'(N)$ are the likelihood curves before and after taking into account the systematic uncertainties;
$\hat{S}$ is the nominal efficiency and $\sigma_{S}$ is its systematic uncertainty. 
As shown in Fig.~\ref{axion_fit}, the combined limits on the relative branching fractions are established at the level of $(3.0-53.0)\times10^{-3}$, and the significance for each case less than $0.5 \sigma$.

\section{SUMMARY}
\begin{figure}[htbp]
\centering
\subfigure{\includegraphics[scale=0.11]{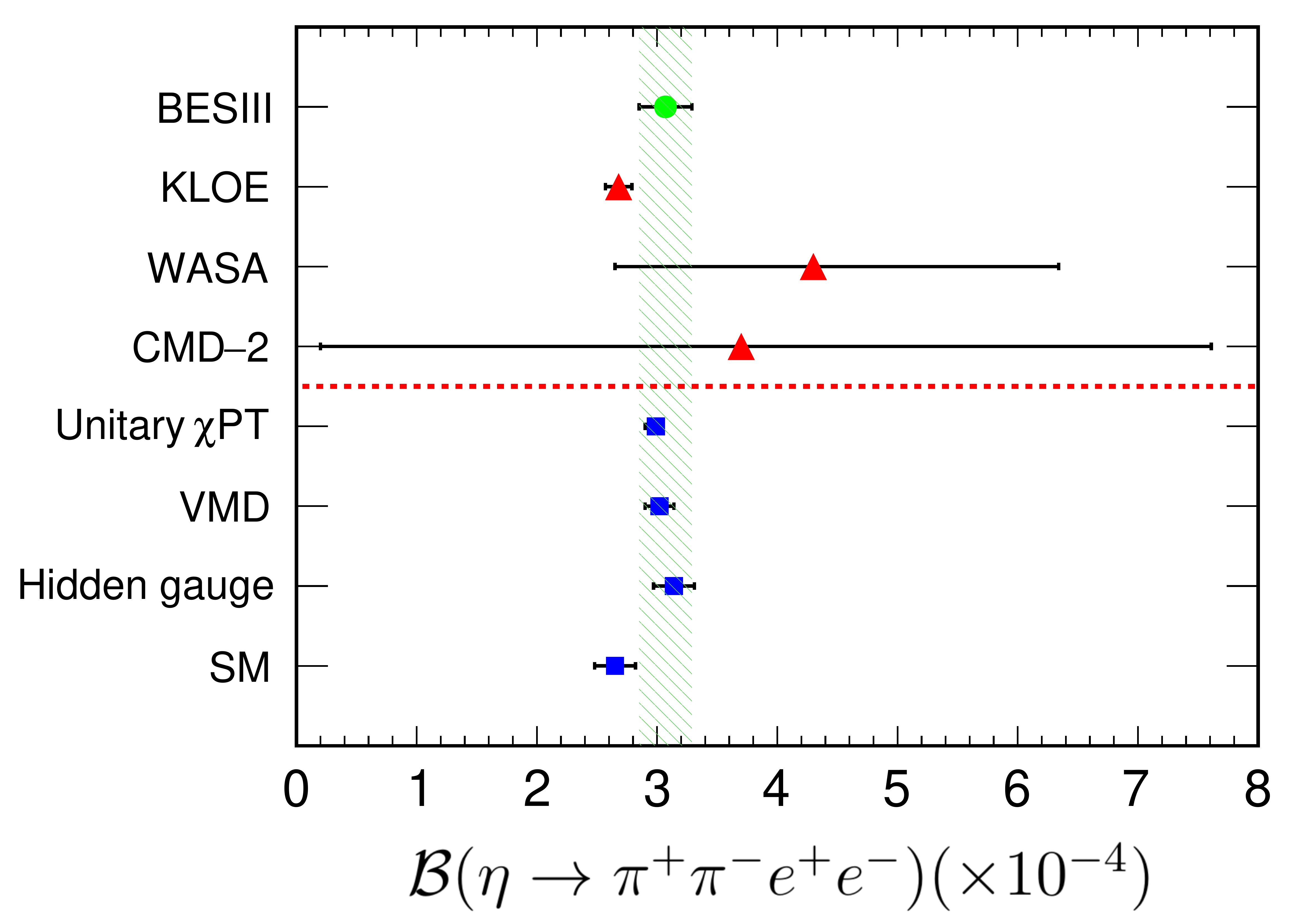}}
\caption{The branching fraction of $\eta\rightarrow\pi^{+}\pi^{-}e^{+}e^{-}$ from different theoretical predictions~\cite{1010_2378, chiral_unitary_approach, Zillinger:2022eva} (blue squares), other experiments~\cite{WASA_2008, CMD2_2001, KLOE_2009} (red triangles) and this measurement (green dot).}
\label{sumcom_br}
\end{figure}

With a sample of $(10087\pm44)\times10^{6}$ $J/\psi$ events, the decays of $\eta\rightarrow\pi^{+}\pi^{-}l^{+}l^{-}$ ($l=e$ or $\mu$) are studied.
For the decay $\eta\rightarrow\pi^{+}\pi^{-}e^{+}e^{-}$, the branching fraction is determined to be $\mathcal{B}(\eta\rightarrow\pi^{+}\pi^{-}e^{+}e^{-})=(3.07\pm0.12_{\rm{stat.}}\pm0.19_{\rm{syst.}})\times10^{-4}$, which is in good agreement with theoretical predictions~\cite{1010_2378, chiral_unitary_approach, Zillinger:2022eva} and previous measurements~\cite{WASA_2008, CMD2_2001, KLOE_2009}, as shown in Fig.~\ref{sumcom_br}.

For the decay $\eta\rightarrow\pi^{+}\pi^{-}\mu^{+}\mu^{-}$, no signal events are observed. The upper limits on the branching fraction are determined to be $\mathcal{B}(\eta\rightarrow\pi^{+}\pi^{-}\mu^{+}\mu^{-})<4.0\times10^{-7}$ at the 90\% C.L., which is improved by 3 orders of magnitude compared to the PDG value $3.6\times10^{-4}$.

Furthermore, the TFF is extracted from the invariant decay amplitude of the reaction $\eta\rightarrow\pi^+\pi^-e^+e^-$. 
An analysis on previous experimental data obtained the values of parameters $c_1 - c_2$ and $c_3$ that are approximately equal to $1$~\cite{ci_parameters_2010} and consistent with Model I (hidden gauge), we thus report our results based on this model. We obtain $m_{V}=(749\pm54_{\rm{stat.}}\pm14_{\rm{syst.}})$ $\mathrm{MeV}/c^{2}$ for the hidden gauge model; a larger $\eta$ data sample is needed to improve precision.

Additionally, the $CP$-violation asymmetry is determined to be $\mathcal{A}_{CP}(\eta\rightarrow\pi^{+}\pi^{-}e^{+}e^{-})=(-4.04\pm4.69_{\rm{stat.}}\pm0.14_{\rm{syst.}})\%$, which implies no $CP$-violation under the present statistics.

Finally, ALPs are searched for via the decay $\eta\rightarrow\pi^+\pi^-a, a\rightarrow e^+e^-$, and the 90\% C.L. upper limits on the branching fraction relative to that of $\eta\rightarrow\pi^+\pi^-e^+e^-$ are presented as a function of the ALP mass within $5-200\ \mathrm{MeV}/c^{2}$, as shown in Fig.~\ref{axion_fit}.
The significance for each case is less than $0.5\sigma$.

\begin{acknowledgments}
The BESIII Collaboration thanks the staff of BEPCII and the IHEP computing center for their strong support. This work is supported in part by National Key R\&D Program of China under Contracts Nos. 2020YFA0406300, 2020YFA0406400, 2023YFA1606000; National Natural Science Foundation of China (NSFC) under Contracts Nos. 11635010, 11735014, 11935015, 11935016, 11935018, 12025502, 12035009, 12035013, 12061131003, 12192260, 12192261, 12192262, 12192263, 12192264, 12192265, 12221005, 12225509, 12235017, 12475089, 12361141819; the Chinese Academy of Sciences (CAS) Large-Scale Scientific Facility Program; the CAS Center for Excellence in Particle Physics (CCEPP); Joint Large-Scale Scientific Facility Funds of the NSFC and CAS under Contract No. U1832207; 100 Talents Program of CAS; The Institute of Nuclear and Particle Physics (INPAC) and Shanghai Key Laboratory for Particle Physics and Cosmology; German Research Foundation DFG under Contracts Nos. 455635585, FOR5327, GRK 2149; Istituto Nazionale di Fisica Nucleare, Italy; Ministry of Development of Turkey under Contract No. DPT2006K-120470; National Research Foundation of Korea under Contract No. NRF-2022R1A2C1092335; National Science and Technology fund of Mongolia; National Science Research and Innovation Fund (NSRF) via the Program Management Unit for Human Resources \& Institutional Development, Research and Innovation of Thailand under Contract No. B16F640076; Polish National Science Centre under Contract No. 2019/35/O/ST2/02907; The Swedish Research Council; U. S. Department of Energy under Contract No. DE-FG02-05ER41374.



\end{acknowledgments}

\bibliography{references}

\end{document}